\documentclass[aps,prl,showpacs,floatfix,twocolumn]{revtex4}
\usepackage{latexsym}
\usepackage{graphicx,subfigure}
\begin{document}
\draft
\thispagestyle{empty}

\title{Analytical solutions for 
black-hole critical behaviour}
\author{$^{1,2}$Tomohiro Harada~\footnote{Electronic
address:harada@rikkyo.ac.jp} and
$^{3}$Ashutosh Mahajan~\footnote{Electronic
address:ashutosh@tifr.res.in}}
\affiliation{$^{1}$Department of Physics, Rikkyo University, Toshima, 
Tokyo 171-8501, Japan,\\
$^{2}$Department of Physics, Kyoto University,
Kyoto 606-8502, Japan, \\
$^{3}$Tata Institute for Fundamental Research, 
Homi Bhabha Road, Mumbai 400005, India}

\date{\today}
\begin{abstract}
Dynamical Einstein cluster is a spherical self-gravitating system of 
counterrotating particles, which may expand, oscillate and collapse.
This system exhibits
critical behaviour in its collapse at the threshold of 
black hole formation.
It appears when the specific angular momentum of particles is 
tuned finely to the critical value.
We find the unique exact self-similar solution at the 
threshold.
This solution begins with a regular surface,
involves timelike naked singularity formation
and asymptotically approaches a static self-similar cluster. 
\end{abstract}
\maketitle

General relativistic numerical simulation (numerical relativity) 
has revealed critical phenomena
at the threshold of black hole formation in self-gravitating 
systems~\cite{choptuik1993}.
When a parameter $p$, which parametrises a generic one-parameter
family of initial data sets, is tuned to the critical value $p^{*}$,
there appears a self-similar solution, which is called a critical
solution. Beyond this value, the collapse ends in a black hole,
its mass $M_{\rm BH}$
obeying the power law $M_{\rm BH}\propto 
|p-p^{*}|^{\gamma}$, where 
$\gamma$ is called a critical exponent. 
The critical behaviour is well described
in terms of the 
behaviour of solutions around a self-similar solution 
with a single unstable mode~\cite{kha1995}.
In this approach, self-similar 
solutions with regularity conditions are numerically found and  
a self-similar solution is numerically shown to be with a single, linearly 
unstable mode.
See~\cite{gundlach2003} for a recent review of 
critical phenomena. 
Apparently, there still is a huge gap between 
numerical simulation and linear stability analysis. 
Moreover, one could suspect unresolved fine structure at the threshold
because almost all results have been based on numerics 
with finite accuracy (cf.~\cite{bsw2004}).

Here we show that there is a system where we can discuss critical phenomena 
in an analytical and exact manner.
This is the spherical system of counterrotating particles,
first introduced by Einstein~\cite{einstein1939} and later generalised
to a dynamical case~\cite{datta1970,bondi1971,evans1976}.
\label{sec:einstein_cluster}
Using a coordinate $r$ comoving to the radial motion of 
each shell, the line element in this
spacetime is given by
\begin{eqnarray}
  ds^2&=&-e^{2\nu(t,r)}dt^2+e^{2\psi(t,r)}dr^2+R^2(t,r)d\Omega^{2}, 
\label{eq:metric}
\end{eqnarray}
where $d\Omega^{2}$ is the line element on the two-dimensional
unit sphere. The Einstein equations and conservation law reduce
\begin{equation}
  \nu^{\prime}=-\frac{1}{h(r,R)}\frac{\partial
    h(r,R)}{\partial R}
  R^{\prime}, 
\quad
  e^{2\psi}=\frac{(R^{\prime})^2 h^2(r,R)}{1+2E(r)}, 
\label{eq:nu'}
\end{equation}
and 
\begin{equation}
  \dot{R}^2 e^{-2\nu}=-1+\frac{2M(r)}{R}+\frac{1+2E(r)}{h^2(r,R)},
  \label{eq:comenergy}
\end{equation}
where $\dot{} \equiv \partial /\partial t$ and $'\equiv \partial /\partial r$,
$M(r)$ and $2E(r)>-1$ are arbitrary functions corresponding to the 
Misner-Sharp mass and the specific energy, respectively,
and $h(r,R)$ is given by 
\begin{equation}
  h^2(r,R)=1+\frac{L^2(r)}{R^2},
\end{equation}
where $L(r)$ is the specific angular momentum
of counterrotating particles.
The energy density is
\begin{equation}
  \epsilon=\frac{M^{\prime}}{4\pi R^2 R^{\prime}}.
    \label{eq:density}
\end{equation}
The motion of each shell is governed by Eq. (\ref{eq:comenergy}) or
\begin{equation}
  \frac{1}{2}\left(\frac{dR}{d\tau}\right)^2+U(r,R)=E(r),
  \label{eq:eom}
\end{equation}
where $d\tau=e^{\nu}dt$ and 
\begin{equation}
  U(r,R)\equiv -\frac{M(r)}{R}+\frac{(1+2E(r))L^2(r)}
			   {2(R^2+L^2(r))}. \label{eq:effpot}
\end{equation}

If we assume that the solution has a regular surface on which all
regular metric functions and physical quantities 
are also analytic, this implies 
the Taylor-series expandability in terms of $R^{2}$.
The arbitrary functions $M$, $E$ and $L^2$ then 
should be expanded as $M=M_3r^3+M_5r^5+\cdots$, 
$ E=E_2r^2+E_4r^4+\cdots$, and $L^{2}=L_{4}^{2}r^4+L_{6}^{2} r^6+\cdots$,
if we choose $r$ so that $r=R$ on the initial regular surface.
$R$ is expanded as $R=R_{1}(t)r+ R_{3}r^3+\cdots$ on a regular surface, 
and using Eq. (\ref{eq:density}), we have 
\begin{equation}
  \epsilon(t,0)=\frac{3M_3}{4\pi R_1(t)^3}.
\label{eq:central_density}
\end{equation}
Observing the lowest order of Eq. (\ref{eq:comenergy}), 
the evolution of $R_1(t)$ is given by
\begin{equation}
  \frac{1}{2}\left(\frac{dR_1}{d\tau}\right)^2=E_{2}+\frac{M_3}{R_1}
-\frac{L_{4}^{2}}{2R_1^2}.
  \label{eq:center}
\end{equation}  
If $L_{4}^{2}>0$, we can find that
$R_1(t)$ necessarily bounces and the neighbouring shell also does
for sufficiently small $r>0$~\cite{evans1976,hin1998}.

\begin{figure}[h]
\includegraphics[width=20pc]{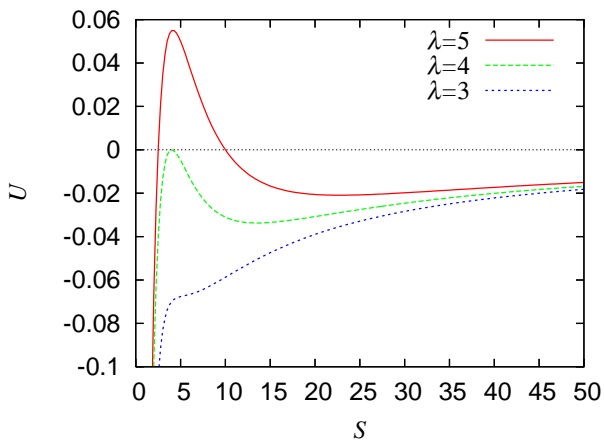}
\caption{\label{fg:potential}
The shape of the potential $U$ for $E=0$ and $L=\lambda M$ with 
$\lambda =3$, 4 and 5. The horizontal axis is $S\equiv R/M$.
}
\end{figure}

The case $L_{4}=0$ was studied in detail in~\cite{hin1998}.
For this case, from Eqs.~(\ref{eq:central_density}) and (\ref{eq:center}), we find that
an initially collapsing cloud inevitably form
a central singularity after a
finite proper time.
As for the motion of the shell for sufficiently small $r>0$,
it turns out that the quantity
$
  \lambda \equiv\lim_{r\to 0}L(r)/M(r)=L_{6}/M_{3}
$
becomes important from Eqs.~(\ref{eq:eom}) and (\ref{eq:effpot}).
Figure~\ref{fg:potential} shows the shape of potentials for different 
values of $\lambda$. For $\lambda >4$, 
the region around $r=0$
necessarily bounces back, while,
for $0\le \lambda <4$, 
the collapse continues to $R=0$.
So, the end state of the collapse
is 
a massless naked singularity or
a black hole of finite mass
depending on the value of $\lambda$.
The critical value $\lambda^{*}$ is $4$.
The mass scaling law and the critical exponent 
will be described elsewhere~\cite{mahajan_etal}.


The self-similarity (homothety) requires that there is a 
vector field ${\bf z}$ such that 
$
{\cal L}_{\bf z}g_{ab}=2g_{ab}.
$
This condition implies 
for the line element~(\ref{eq:metric}) that
there is a coordinate system $(t,r)$ such that 
$\nu$, $\psi$ and $S\equiv R/r$ are
functions of $z\equiv t/r$~\cite{ct1971}.
For dynamical Einstein clusters, this means 
$E(r)=E$, $M(r)=\mu r$ and $L(r)=l r$, where $E$, $\mu$ and $l$
are constants.
Then Eqs.~(\ref{eq:nu'}) and (\ref{eq:comenergy}) 
reduce the following closed autonomous system:
\begin{eqnarray}
\frac{d\sigma}{d\ln|z|}-1&=&\frac{l^{2}}{S^{2}+l^{2}}
\frac{1}{S}\left(\frac{d S}{d\ln |z|}-S\right), 
\label{eq:dnudz}\\
e^{-2\sigma}\left(\frac{dS}{d\ln|z|}\right)^{2}&=&
-1+\frac{2\mu}{S}+\frac{(1+2E) S^{2}}{S^{2}+l^{2}},
\label{eq:dsdz}
\end{eqnarray}
where $\sigma\equiv \nu+\ln |z|$.
We can eliminate $z$ from the above equations
and get
\begin{equation}
\frac{de^{\sigma} }{d S}=
\frac{l^{2}}{S(S^{2}+l^{2})}e^{\sigma}
\pm \frac{S^{2}}{S^{2}+l^{2}}\frac{1}
{\sqrt{2(E-U(S))}},
\label{eq:dsigmads}
\end{equation}
where the upper (lower) sign corresponds to the positive (negative)
sign of $dS/d (\ln |z|)$ and
\begin{eqnarray}
U(S)&\equiv & -\frac{\mu}{S}+\frac{1}{2}\frac{(1+2E) l^{2}}{S^{2}+l^{2}}.
\end{eqnarray}
Equation~(\ref{eq:dsdz}) reduces
\begin{equation}
\frac{1}{2}\left(\frac{dS}{d\xi}\right)^{2}+U(S)=E,
\label{eq:eomss}
\end{equation}
where $d\xi^{2} \equiv 
e^{2\nu(z)}dz^2=e^{2\sigma}(d\ln |z|)^{2}$,
$\xi$ corresponding to the nondimensional proper time for 
the comoving observer.
If we use $S$ as a coordinate in place of $t$, we get
\begin{eqnarray}
ds^{2}&=&e^{2\zeta}\left[-\left(e^{\sigma} d\zeta 
\pm \frac{dS}{\sqrt{2(E-U(S))}}\right)^{2}
+ \frac{S^{2}+l^{2}}{(1+2E) S^{2}} 
\right. \nonumber  \\
& \times& \left.
\left(S\mp e^{\sigma} \sqrt{2(E-U(S))}\right)^{2}d\zeta^{2}
+S^{2}d\Omega^{2}\right],
\label{eq:metric_rs}
\end{eqnarray}
where $\zeta \equiv \ln r$ and $S$ plays a role of similarity variable.
From Eq.~(\ref{eq:density}), 
\begin{equation}
4\pi \epsilon r^{2}=
\frac{\mu}{S^{2}\left(S\mp e^{\sigma}\sqrt{2(E-U(S))}\right)}.
\label{eq:4pirhor2}
\end{equation}
If $R'=0$ but $R\ne 0$, there appears shell-crossing
	   singularity and the model gets invalid beyond there.
This condition reduces the following equation:
\begin{equation}
S\mp e^{\sigma}\sqrt{2(E-U(S))}=0.
\label{eq:shell-crossing}
\end{equation}
The solution is shell-crossing free where $dS/d\ln|z|<0$.

When we rescale $t$ and $r$ as $\tilde{t}=at$ and $\tilde{r}=ar$, 
we will get $\tilde{E}=E$, $\tilde{\mu}=\mu/a$, 
$\tilde{l}=l/a$, 
$\tilde{\sigma}=\sigma-\ln a$ and $\tilde{S}=S/a$ and the above
equations are invariant.
Hereafter, we fix this scaling so that $\mu=1$ and $l=\lambda$, 
in which $r$ coincides with the mass and is denoted as $m$.
Note that we still have the freedom of rescaling as $\tilde{t}=at$
because this only changes $z$ to $\tilde{z}=az$ and 
Eqs.~(\ref{eq:dnudz})--(\ref{eq:dsdz}) are invariant.
Except for this gauge freedom, 
there are a three-parameter family of self-similar solutions
parametrised by $E$, $\lambda$ and the initial value for
$\sigma=\sigma_{i}$ at $S=S_{i}$.

At the regular centre, $m/R$ vanishes or $S=\infty$.  
To assure the circular constant being $\pi$, $h^{2}/(1+2E)\to 1$ 
as $S\to \infty$, implying $E=0$. Moreover, we assume
that the cluster has a critical angular momentum, i.e., the right-hand 
side of Eq.~(\ref{eq:dsdz}) has a double root, 
implying $\lambda =4$.
Then, we have specified two parameters and 
have a one-parameter family of self-similar solutions.
For $E=0$ and $\lambda=4 $,
the shape of the potential is shown in Fig.~\ref{fg:potential}.
Equation~(\ref{eq:eomss}) then 
has essentially three solutions
and all others are generated through time reversal or time translation
of those solutions. 
The first one is a static solution at the top of the potential.
For this case, we assume $\psi$ and $R$ 
have no dependence on $t$ but do not for $\nu$.
This is because $\partial/\partial t$ may not coincide
with a timelike Killing vector.
Then, Eqs.~(\ref{eq:dnudz})--(\ref{eq:dsdz}) yield the following
solution:
$e^{2\sigma}=c_{0}^{2}|z|$, 
$e^{2\psi}= 32$, and
$S=4$,
where $c_{0}$ is a constant and set to be unity, using the rescaling of
$t$. The resulting metric is given by
\[
ds^{2}
=-R  dT^{2}+2 dR ^{2}+R ^{2}d\Omega^{2},
\]
where we have implemented the coordinate transformation:
$dt^{2}/|t|=4 dT^{2}$ and $4 m=R $.
The density $\epsilon$ is given by
$
4\pi \epsilon=1/(64 m^{2})=1/(4 R ^{2}).
$
This spacetime has a timelike naked singularity at the centre and 
suffers from a solid angle deficit so that
the area of the sphere $R=\mbox{const}$ divided by the squared radius
is not $4\pi$ but $2\pi$.
It can be easily shown that static singular 
solutions, which were discovered in~\cite{khi2000}, 
all fall into this 
self-similar static solution after an 
appropriate coordinate transformation.


The remaining two solutions are dynamical.
The second one is the solution,
in which $S$ begins with infinity 
at $\xi=-\infty$, monotonically decreases and asymptotes 4
as $\xi$ increases to $\infty$.
The third one is the solution, in which 
$S$ begins with 4 at $\xi=-\infty$, monotonically 
decreases to 0 at a finite value of $\xi$.
For these two solutions, we can express $e^{\sigma}$ in terms of $S$ using 
only elementary functions as an integral of Eq.~(\ref{eq:dsigmads}):
\begin{equation}
e^{2\sigma}=\frac{S^{2}}{S^{2}+16}
\left(\frac{\sqrt{2S}}{3}(S+12)
+4\sqrt{2}\ln \left|\frac{\sqrt{S}-2}{\sqrt{S}+2}
\right|+A \right)^{2},
\label{eq:integral}
\end{equation}
where $A$ is an arbitrary constant.
This is actually a subclass of solutions whose expressions 
were given in terms of 
elementary functions in~\cite{hin1998}.
Only the second solution
has a regular centre because $m/R$ is always equal to and larger than
a quarter for the static solution and the third solution,
respectively.

Let us concentrate on the second solution.
Around the regular centre $S= \infty$, from Eq.~(\ref{eq:integral}), we have
\begin{equation}
e^{\sigma}= \frac{\sqrt{2}}{3}S^{3/2}+4\sqrt{2}S^{1/2}+A+O(S^{-1/2}).
\end{equation}
Then, from Eqs.~(\ref{eq:dnudz}) and (\ref{eq:dsdz}),
we have for the lowest order term 
$
S\approx  (9/2)^{1/3}X^{2/3}|z|^{2/3} $ and 
$ e^{\sigma}\approx X|z|$, 
as $|z| \to \infty$, where the constant $X$ can be
set to be unity by rescaling $t$.
This is exactly the behaviour of the regular centre:
$R \propto m^{1/3}$ and $e^{\nu} \to 1 $ at $m\to 0$.
In this regime, 
$
\xi= z+\mbox{const},
$
and $t$ coincides with the proper time at the regular centre. 
Equation~(\ref{eq:4pirhor2}) implies
$
4\pi \epsilon_{0}(t) t^{2}=2/3,
$
where $\epsilon_{0}(t)$ is the density at the regular centre.
The central density diverges
to infinity at $t=0$, resulting in a central singularity.
In other words, $t=0$ is characterised with the appearance
of the central singularity.
Let us assume the expandability for the density
on the regular surface $t=\mbox{const}<0$:
\begin{equation}
\epsilon=\sum_{i=0}^{\infty}\epsilon_{i}(t)R^{i}.
\label{eq:density_expansion}
\end{equation}
On this surface, we can write down $S=R/m$ as
\begin{equation}
S=\left(\frac{4}{3}\pi \epsilon_{0}R^{2}\right)^{-1}
\left(1+\frac{3}{\epsilon_{0}}\sum_{i=1}^{\infty}\frac{\epsilon_{i}}{i+3}R^{i}
\right)^{-1}.
\label{eq:S_expansion}
\end{equation}
Substituting Eqs.~(\ref{eq:density_expansion}) and (\ref{eq:S_expansion})
into Eq.~(\ref{eq:4pirhor2}) and comparing both sides, we get
$\epsilon_{1}=0$, $\epsilon_{2}=(160/3)\pi \epsilon_{0}^{2}$ and
$\epsilon_{3}=16\sqrt{2/3}\pi^{3/2}A\epsilon_{0}^{5/2}$.
Therefore, the density around the regular centre can be expanded as
\begin{equation}
\epsilon=\frac{1}{6\pi t^{2}}+\frac{40}{27\pi}\frac{R^{2}}{t^{4}}+
\frac{4}{27\pi}A\frac{R^3}{|t|^{5}}+O(R^{4}).
\end{equation}
The density takes a local minimum at the centre.
If we assume analyticity on the regular surface,
$A=0$ is concluded and there no longer appear 
odd power terms of $R$ in the expansion.
This analyticity requirement has 
been imposed for the identification of critical solutions~\cite{kha1995}.
So we will identify the second solution for $A=0$ with the unique
threshold solution.
Note that the analysis below is nevertheless also 
applicable to the case $A\ne 0$.

Let us see other physical properties of this solution.
We can find from Eq.~(\ref{eq:integral}) that $e^{\sigma}$
can be zero. Let $\alpha(>4)$ such that $e^{\sigma}=0$ at $S=\alpha$. 
$\alpha (=4.8133\cdots)$ is a root of the following transcendental equation:
\[
\sqrt{S}(S+12)=12\ln\frac{\sqrt{S}+2}{\sqrt{S}-2}.
\]
Around there, from Eqs.~(\ref{eq:dsdz}) and (\ref{eq:dsigmads}), 
we can find
$
|S-\alpha|\propto |z|^{\beta}$
and 
$e^{\sigma}\propto |z|^{\beta}$
as $z\to 0$, where $\beta=\alpha^{2}/(\alpha^{2}+16)$ and hence
$1/2<\beta<1$. 
In this regime, 
$
\xi=\mp B |z|^{\beta} + \mbox{const},
$
where $B$ is a constant.
This means that it takes only a finite proper 
time to reach $S=\alpha$ or $t=0$. 
Thus, $t=0$ is only a coordinate singularity and the solution can be 
extended regularly and uniquely beyond $t=0$.
This is clear in Eq.~(\ref{eq:metric_rs}), where 
there is no singularity at $S=\alpha$.
The density profile at $t=0$ follows an exact power law
$
4\pi \epsilon =1/(\alpha R^{2}).
$
Because of self-similarity, $z=0$ also corresponds to 
infinity ($|t|<\infty$ and $m=\infty$).
This really corresponds to the surface of infinite area $R=\infty$.
It follows that the density falls off as
$
4\pi \epsilon \approx 1/(\alpha R^{2})
$
as $R\to \infty$. This behaviour is time-independent.

Each shell dynamically approaches the specific radius
$S=4$ as $\xi\to \infty$.
For $S\to 4$, from Eq.~(\ref{eq:integral}), we have
$
e^{\sigma} \approx -4 \ln (S-4).
$
Then, from Eqs.~(\ref{eq:dnudz}) and (\ref{eq:dsdz}),
we have 
$(S-4)\approx e^{-C |z|^{1/2}}$ and $e^{\sigma}=4C|z|^{1/2}$
as $|z| \to \infty$, where $C>0$ is a constant. 
This means that the asymptotic solution
is the static solution.
In this regime, 
$
\xi\approx 8B |z|^{1/2} + \mbox{const},
$
and 
$
\tau\approx - 8 m \ln(S-4).
$
Hence the collapse approaches the static solution 
with an infinite proper time. 

Since $dS/d\ln|z|>0$ for $t<0$ for the threshold solution, 
we need to check whether there is a root in $(\alpha ,\infty)$ 
of Eq.~(\ref{eq:shell-crossing}) or
the following equation:
\[
\frac{2(S-4)}{\sqrt{S}(S^{2}+16)}
\left|\frac{\sqrt{S}}{3}(S+12)
+4\ln \frac{\sqrt{S}-2}{\sqrt{S}+2} \right|=1.
\]
When $S$ increases from $\alpha$ to $\infty$, the left-hand side
increases from 0, takes a maximum and decreases to 2/3.
This maximum value is $0.76412\cdots$, 
well below unity. 
Since $dS/d\ln|z|<0$ for $t>0$,
the threshold solution is free of shell-crossing.

\begin{figure}[h]
\begin{tabular}{cc}
\includegraphics[width=20pc]{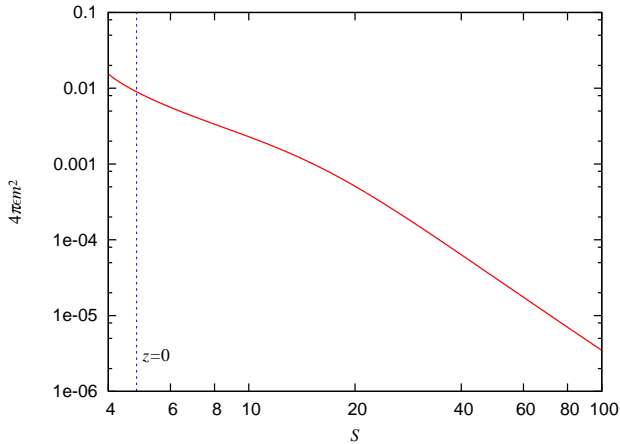}
\end{tabular}
\caption{\label{fg:rho}
The nondimensional density $4\pi \epsilon m^{2}$ in terms of $S$
for the exact threshold solution.
}
\end{figure}

\begin{figure}[h]
\begin{tabular}{cc}
\includegraphics[width=20pc]{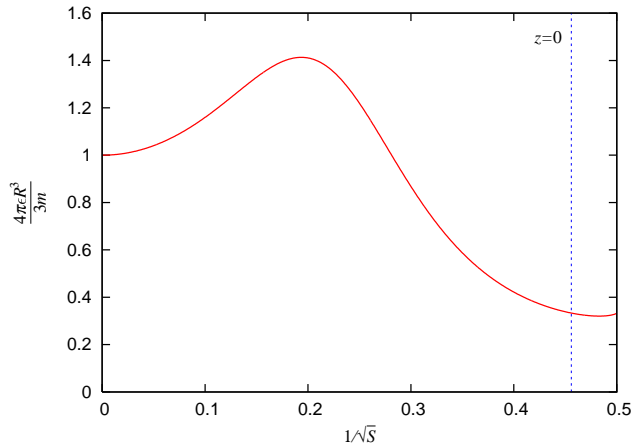}
\end{tabular}
\caption{\label{fg:normalised_density}
$(4\pi\epsilon R^3)/(3m)$, the ratio of the local density 
to the averaged density interior to the 
shell in terms of $1/\sqrt{S}$ 
for the exact threshold solution.}
\end{figure}

Figure~\ref{fg:rho} shows 
the evolution of the density in terms of $S$ from 
Eq.~(\ref{eq:4pirhor2}).
As time proceeds from $t=-\infty$ to $t= \infty$, 
$S$ monotonically decreases from $\infty$ to 4
and the density $\epsilon$ observed by a comoving observer 
monotonically increases from $0$ to the value for the 
static solution $1/(256\pi m^{2})$. It smoothly crosses $z=0$ 
at $S=\alpha$ and $\epsilon =1/(4\pi \alpha^3 m^{2})$.
Figure~\ref{fg:normalised_density} shows the ratio of 
the local density $\epsilon$ 
to the averaged density inside the shell $m/(4\pi R^3/3)$ ,
in which the horizontal axis is $1/\sqrt{S}$.
This ratio must be unity at the regular centre and 
$1/\sqrt{S}$ is proportional to the area radius around the regular centre.
We can see that the density takes a local minimum at the regular
centre $S=\infty$ or $z=-\infty$, 
increases around the centre, takes a maximum and decreases outside.
At spatial infinity $S=\alpha$ or $z=0$, 
this ratio becomes $1/3$, implying that 
$\epsilon$ falls off in proportion to $R^{-2}$. The ratio goes below $1/3$
as $z$ increases further and comes back to $1/3$ at $S=4$ or 
$z=\infty$, implying that $\epsilon$ again gets
proportional to $R^{-2}$ as $t\to\infty$.

There is no trapped surface because $S=R/m>4$ is
satisfied everywhere. So, the central singularity is naked.
In fact, this spacetime is a member of solutions for which 
the causal structure is shown in Figure 1 of~\cite{khi2000}.
The central singularity is naked and timelike.

The general solutions of the dynamical Einstein cluster 
is exactly solved in terms of elliptic
integrals using the mass-area coordinates~\cite{magli1997,hin1998}
and here the critical self-similar solution is uniquely
obtained in terms of elementary functions.
This system provides a tractable laboratory for studying how 
generic critical collapse approaches the 
threshold solution in both linear and 
nonlinear regimes. 

\acknowledgments
TH is very grateful to K.~Nakao and H.~Iguchi for helpful comments.
This work was partly supported by the
Grant-in-Aid for the 21st Century
COE ``Center for Diversity and Universality in Physics''
and the Grant-in-Aid for Scientific Research Fund 
(Young Scientists (B) 18740144)
of the Ministry of Education, Culture, Sports, Science and Technology
of Japan.

\end{document}